\documentclass[pra,showkeys,showpacs,groupedaddress,twocolumn]{revtex4}
\usepackage{amssymb}
\usepackage{graphicx}
\usepackage{dcolumn}
\usepackage{amssymb}
\usepackage{bm}
\usepackage{amsmath}
\usepackage{subfigure}
\usepackage{float}
\usepackage{color}
\usepackage{indentfirst}
\usepackage{array}
\usepackage{multirow}
\usepackage{supertabular}
\usepackage{longtable}
\usepackage{epsfig,psfrag}

\begin{document}
\title{Nonlinear Zeeman effect, line shapes and optical pumping in electromagnetically induced transparency}
\author{Linjie Zhang$^{1,2}$}
\thanks{Corresponding author: zlj@sxu.edu.cn}
\author{Shanxia Bao$^{1,2}$}
\author{Hao Zhang$^{1,2}$}
\author{Georg Raithel$^{1,3}$}
\thanks{Corresponding author: graithel@umich.edu}
\author{Jianming Zhao$^{1,2}$}
\author{Liantuan Xiao$^{1,2}$}
\author{Suotang Jia$^{1,2}$}
\affiliation{$^{1}$State Key Laboratory of Quantum Optics and Quantum Optics Devices, Institute of Laser Spectroscopy, Shanxi University, Taiyuan 030006, China}
\affiliation{$^{2}$Collaborative Innovation Center of Extreme Optics, Shanxi University, Taiyuan 030006, China}
\affiliation{$^{3}$Department of Physics, University of Michigan, Ann Arbor, MI 48109-1120, USA}

\begin{abstract}
We perform Zeeman spectroscopy on a Rydberg electromagnetically induced transparency (EIT) system in a room-temperature Cs vapor cell, in magnetic fields up to 50~Gauss and for several polarization configurations. The magnetic interactions of the $\vert 6S_{1/2}, F_g=4 \rangle$ ground,  $\vert 6P_{3/2}, F_e=5 \rangle$ intermediate, and $\vert 33S_{1/2} \rangle$ Rydberg states that form the ladder-type EIT system are in the linear Zeeman, quadratic Zeeman, and the deep hyperfine Paschen-Back regimes, respectively. Starting in magnetic fields of about 5~Gauss, the spectra develop an asymmetry that becomes paramount in fields $\gtrsim40$~Gauss. We use a quantum Monte Carlo wave-function approach to quantitatively model the spectra. Simulated spectra are in good agreement with experimental data. The asymmetry in the spectra is, in part, due to level shifts caused by the quadratic Zeeman effect, but it also reflects the complicated interplay between optical pumping and EIT in the magnetic field. Relevance to measurement applications is discussed.
%The simulations are also used to study optical pumping in the magnetic field and to investigate the interplay between optical pumping and EIT, which reduces photon scattering and optical pumping.
\end{abstract}
\pacs{32.60.+i,32.80.Xx,32.80.Ee, 42.50.Gy}
\keywords{nonlinear Zeeman effect,optical pumping,Rydberg EIT,quantum Monte Carlo wave-function}
\maketitle

\section{Introduction}

Rydberg atoms, highly excited atoms with principal quantum numbers $n \gg$1 ~{\color{blue}\cite{Gallagher1}}, possess extraordinary properties, such as long lifetime and strong dipole-dipole interaction. These offer considerable potential for applications in quantum information processing~{\color{blue}\cite{Jaksch2, Urban3}}, optical non-linearity~{\color{blue}\cite{ Dudin4, Peyronel5}} and non-equilibrium phenomena~{\color{blue}\cite{Diehl6, Carr7, Marcuzzi8}}. Their strong response to electric fields (DC polarizabilities typically scale $\propto$ $n^{7}$) makes these atoms attractive for field measurement purposes~{\color{blue}\cite{Zimmerman9, Sedlacek10}}. Electromagnetically induced transparency (EIT)~{\color{blue}\cite{Harris11}}, a quantum interference effect, has been extended to ladder-type systems that include Rydberg levels. A. K. Mohapatra et al. introduced EIT as an optical, nondestructive method to probe Rydberg states in vapor cells~{\color{blue}\cite{Mohapatra12}}. Later, a giant electro-optic effect based on polarizable dark states was observed that enables precision electrometry~{\color{blue}\cite{Mohapatra13}}. D. S. Ding et al. studied optically-driven non-equilibrium phase transitions with a much higher sensitivity by using Rydberg EIT in a thermal cell~{\color{blue}\cite{Ding14}}. We have demonstrated the important role of the laser-field polarizations in Rydberg EIT, and qualitatively modeled its Zeeman spectra using standard density matrix method~{\color{blue}\cite{Bao15}}.

Recently, it has been shown that the Zeeman effect and optical pumping play important roles in quantum information processing and Rydberg-atom molecules~{\color{blue}~\cite{Urban3,Bendkowsky16,Bottcher17}}. Level populations and coherences in ladder-type Rydberg EIT systems
are affected by the Zeeman shifts of the magnetic sub-states of the ground, intermediate and Rydberg levels. The systems exhibit an interdependence between photon scattering rates and optical pumping,  linear and quadratic Zeeman splittings, and EIT (which reduces photon scattering). In vapor cells, due to the Doppler mismatch between probe and coupling lasers~{\color{blue}\cite{Xiao18}}, a single spectrum exhibits multiple EIT peaks, each of which is associated with EIT on different sequences of Zeeman sub-states. Any use of Rydberg-EIT in electric-field diagnostics under presence of a magnetic field requires an understanding of the detailed EIT line shapes in the magnetic field. The line shapes depend on level shifts and optical pumping. The latter can be greatly reduced by Zeeman detunings, with asymmetries following from the quadratic Zeeman effect. 

In this paper, we study Rydberg EIT in a vapor cell in a magnetic field that is parallel with the laser-beam directions. The Rydberg EIT spectra exhibit quadratic Zeeman shifts and asymmetries that develop with increasing magnetic field, and that become very significant at fields as low as a few tens of Gauss. Our study shows that these features are caused by the quadratic Zeeman effect of the 6$\emph{P}_{3/2}  F_{e} = 5$ levels, and by the interplay between EIT and optical pumping. We employ a quantum Monte-Carlo wave-function (QMCWF) approach to model Zeeman spectra, optical pumping, and asymmetry of Rydberg-EIT lines in fields up to 50~Gauss. Simulation results are in good agreement with our experiments and afford insight into the optical-pumping dynamics.

\section{Experimental setup} 

The ladder three-level system is formed by the ground state 6$\emph{S}_{1/2} F_g=4$, the intermediate state 6$\emph{P}_{3/2} F_e=5$ and the 33$\emph{S}_{1/2}$ Rydberg state of $^{133}$Cs, denoted $|g\rangle$, $|e\rangle$ and $|r\rangle$, respectively, as shown in Fig.~1(a)(left). The experiments are performed in a room-temperature Cs vapor cell with the counter-propagated probe and coupling beams focused into the center of the cell. The experimental setup is sketched in Fig.~1(b). The laser beams propagate in $z$-direction (quantization axis), paralleled to the magnetic field. The probe laser (DL100, Toptica) has a wavelength $\lambda_P=852~$nm. The probe beam has a power of 2~$\mu$W and a 1/$e^{2}$ waist of $w_0=130~\mu$m. The probe laser is locked on a high-finesse Fabry-Perot cavity with a spacer made from ultra-low thermal expansion glass (Stable Laser Systems ATF-6010-4) and has a linewidth smaller than 100~kHz. The probe laser resonantly drives the 6$\emph{S}_{1/2} F_g=4 \rightarrow$ 6$\emph{P}_{3/2} F_e=5$ transition. The coupling laser is a frequency-doubled Toptica TA-SHG Pro system and has a wavelength $\lambda_C=510$~nm. The coupling beam has a power of 20~mW, a 1/$e^{2}$ waist of 130~$\mu$m, and a linewidth of about 1~MHz. The coupling laser is scanned through the 6$\emph{P}_{3/2} F_e=5 \rightarrow 33\emph{S}_{1/2}$ Rydberg-state transition. The probe laser is locked on-resonance, and its transmission through the sample is monitored while the coupling laser is scanned. In this method, there is no variation of the coupling-beam-free absorption signal, as the probe always resonantly interacts with the same velocity classes in the Maxwell velocity distribution. The coupling-beam-induced EIT transmission peaks then appear on a flat background.

As sketched in Fig.~1(b), the Cs vapor cell (length 4~cm and diameter 2~cm) is contained in a much longer cylindrical solenoid (length 25~cm, inner diameter 5~cm, three layers of 211 windings each), so that the uniformity of the magnetic field along the length of the vapor cell is guaranteed. The magnetic field is set to values between 0 and 50~Gauss, with a variation of less than $\pm 0.1$~Gauss within the cell. The solenoid is enclosed within several layers of magnetic-shielding material to eliminate the influence of environmental magnetic fields. To ensure
clean polarizations, both the coupler and probe beams are passed through combinations of half-wave plates ($\lambda$/2), Glan-Taylor polarizers (GTP) and quarter-wave plates ($\lambda$/4) located immediately before the beams enter the cell.

We use an auxiliary magnetic-field-free EIT reference setup (not shown), which is similar to the one sketched in Fig.~1(b) and is operated with the same lasers as the main setup. When scanning the coupler laser, the EIT spectrum from the reference cell exhibits peak pairs that mark the magnetic-field-free  6$\emph{S}_{1/2}, F_e=4$ $\rightarrow$ $6\emph{P}_{3/2} F_e=5 \rightarrow$ 33$\emph{S}_{1/2}$ and 6$\emph{S}_{1/2} F_g=4 \rightarrow$ 6$\emph{P}_{3/2} F_e=4 \rightarrow$ 33$\emph{S}_{1/2}$ cascade transitions.
The coupler frequency at which the 6$\emph{S}_{1/2}, F_e=4$ $\rightarrow$ $6\emph{P}_{3/2} F_e=5 \rightarrow$ 33$\emph{S}_{1/2}$
reference peak occurs defines $\Delta_C = 0$ in the EIT spectra obtained with the main setup.
Meanwhile, the separation between the two reference peaks is 168~MHz ({\sl{i. e.}}, the actual $F_e=4$ to $F_e=5$ splitting multiplied with a Doppler correction factor $(\lambda_P/\lambda_C)-1$). This allows us to calibrate the coupler-frequency scans~{\color{blue}\cite{Jiao19}}. The calibrations are essential for precise measurement of the absolute Zeeman splittings in an applied external magnetic field.

\begin{figure}[h]
\centering
\includegraphics[width=0.45\textwidth]{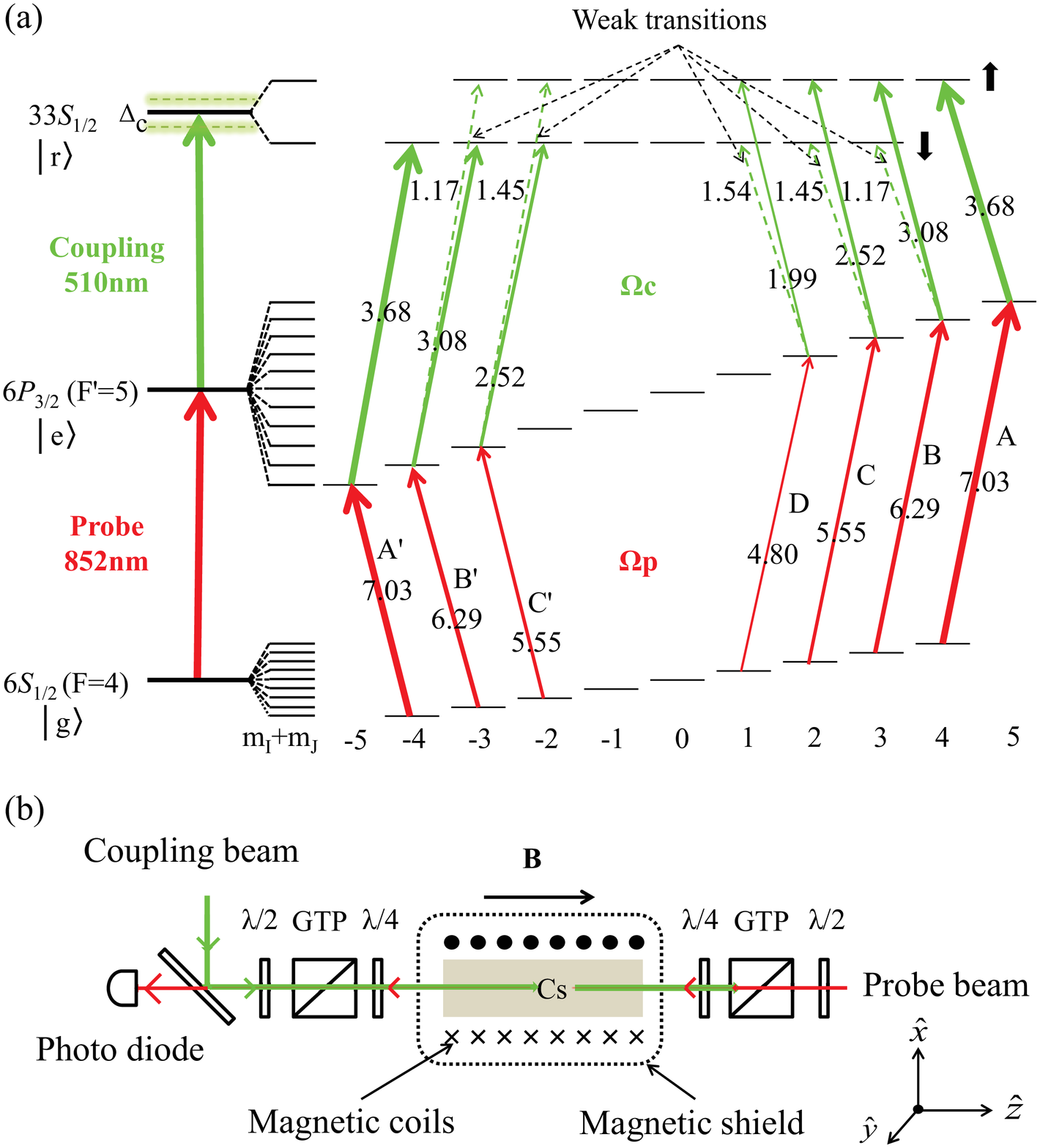}
\caption{(Color online) (a) Energy-level scheme of Rydberg EIT without (left) and with (right) magnetic field. The 852-nm probe laser is resonant with the field-free transition 6$\emph{S}_{1/2} F_g=4 \rightarrow$ 6$\emph{P}_{3/2} F_e=5$, and the 510-nm coupling laser scans through the 6$\emph{P}_{3/2} F_e=5 \rightarrow$ 33$\emph{S}_{1/2}$ Rydberg-state transition. The coupler detuning relative to the field-free transition 6$\emph{P}_{3/2} F_e=5 \rightarrow$ 33$\emph{S}_{1/2}$ is denoted $\Delta_{c}$. The sketch on the right shows the Zeeman sublevels of the $|g\rangle$, $|e\rangle$ and $|r\rangle$ states when a magnetic field on the order of ten Gauss is applied (main separations not to scale). The level ladders that lead to strong signals in the EIT spectra measured with $x$-polarized fields are labeled A, B, C, D and A$'$, B$'$, C$'$ (line thickness increases with line strength). The numbers show Rabi frequencies $\Omega$/2$\pi$ in MHz.
The stronger coupling transitions (solid green lines, Type I) approximately maintain the electron spin, while
in the weaker ones (dashed green lines, Type II) the spin mostly flips.  (b) Schematic of the experimental setup. The probe and coupling beams are counter-propagated and focused into the center of cell, which is contained in a cylindrical solenoid and multi-layer shield. The indicated coordinates define magnetic-field and polarization directions. Half-wave plates ($\lambda$/2), Glan-Taylor polarizers (GTP) and quarter-wave plates ($\lambda$/4) are used for polarization control.}
\label{fig1}
\end{figure}

\section{Theoretical methods}

\subsection{Hyperfine Hamiltonian and EIT resonances}

In our calculations we employ the $\{ |m_I, m_J \rangle \}$ basis, with $m_I = -3.5, -2.5, ..., 3.5$ for $^{133}$Cs and $m_J = \pm 0.5$ for $\vert g \rangle$ and $\vert r \rangle$, and $m_J = \pm 0.5, \pm 1.5$ for $\vert e \rangle$. In this basis, the computation of the matrix elements of the below operators is straightforward.

For an atom with velocity $v$ in $z$-direction, the Doppler-corrected field-free part  of the total effective Hamiltonian is, in the field picture,
\begin{eqnarray}
\hat{H}_{0} & = & -\hbar (\Delta_P + k_P v + \frac{{\rm{i}}}{2} \Gamma_e ) \vert e \rangle \langle e \vert \nonumber\\
         ~  & ~ & -\hbar (\Delta_P + \Delta_C + (k_P-k_C) v + \frac{{\rm{i}}}{2} \hbar \Gamma_r ) \vert r \rangle \langle r \vert
\label{eq:h0}
\end{eqnarray}
with probe and coupling-laser detunings $\Delta_P$ (fixed and $\approx 0$) and $\Delta_C$ (scanned), wavenumbers $k_i$ for $i=P,C$, intermediate-state decay rate $\Gamma_e = 2 \pi \times 5.2~$MHz, and short-hands for the intermediate- and Rydberg-state projectors $\vert e \rangle \langle e \vert$ and $\vert r \rangle \langle r \vert$ (which include sums over the magnetic quantum numbers). The Rydberg-state decay rate $\Gamma_r$ is negligibly small (we usually set it to $2 \pi \times 10~$kHz). When solving the time-dependent Schr\"odinger equation, the imaginary parts of the effective Hamiltonian lead to an exponential decay of the norm of the wavefunction that is consistent with the decay rates and wave-function probabilities in $ \vert e \rangle$ and $\vert r \rangle$. The wavefunction decay is important in the implementation of the quantum Monte-Carlo wave-function (QMCWF) method.

The hyperfine Hamiltonian is given by
\begin{flalign}
&\hat{H}_{hfs}=A_{hfs}\hat{\textbf{I}}\cdot\hat{\textbf{J}}&\nonumber\\
& +B_{hfs}\frac{[3(\hat{\textbf{I}}\cdot\hat{\textbf{J}})^{2}
 +3/2(\hat{\textbf{I}}\cdot\hat{\textbf{J}})-I(I+1)J(J+1)]}{2I(2I-1)J(2J-1)}&
\label{eq:hhfs}
\end{flalign}
where $A_{hfs}$, $B_{hfs}$ are the magnetic dipole and electric quadrupole constants of the hyperfine structure (HFS), respectively (the magnetic octupole interaction is neglected). Values for the ground- and intermediate-state hyperfine constants are taken from Ref.~{\color{blue}~\cite{Steck20}}. The electronic- and nuclear-spin operators are denoted $\hat{\textbf{J}}$
and $\hat{\textbf{I}}$, respectively.
For the ground and Rydberg states $B_{hfs}=0$. For $nS_{1/2}$ Rydberg levels $A_{hfs}=
13.5~$GHz$/(n-\delta_S)^3$~{\color{blue}~\cite{Sassmannshausen21}}, with quantum defect $\delta_S=4.05$. For $33S_{1/2}$, the Rydberg state used in this work, the Rydberg HFS is negligible at the present level of precision.

For a magnetic field $B$ pointing in $z$-direction, the magnetic interaction is given by
\begin{flalign}
&\hat{H}_{B}=\frac{\mu_{B}}{\hbar}(g_{J}\hat{\textbf{J}}+g_{I}\hat{\textbf{I}})\cdot\textbf{B}=
\frac{\mu_{B}}{\hbar}(g_{J}\hat{J}_{z}+g_{I}\hat{I}_{z})B&
\label{eq:hB}
\end{flalign}
where $\mu_{B}/h$=1.4 MHz/Gauss is the Bohr magneton, and $g_{J}$ and $g_{I}$ are the Lande factor and the nuclear g-factor ($g_{I} \ll g_{J}$).

The optical atom-field interactions are, in the field picture,
\begin{equation}
\hat{H}_{i}= E_i \epsilon_i \cdot \hat{\textbf{d}}_i/2
\label{eq:hi}
\end{equation}
with field amplitudes $E_i$, polarization unit vectors $\epsilon_i$ and electric-dipole operators  $\hat{\textbf{d}}_i$, for the probe and coupler fields $i=P$ and $C$, respectively. In the geometry in Fig.~1, we employ polarization unit vectors $\epsilon_i = \hat{x}$ or $\epsilon_i = (\hat{x} \pm {\rm{i}} \hat{y})/\sqrt{2}$, for $x$- and circular polarizations, respectively. The probe-laser interaction matrix elements are of the  form  $\hbar \Omega_P (m'_J,m_J)/2 = \hbar \Omega_{P,r} \left[ (w)^{l'=1,J'=3/2,m'_J}_{l=0,J=1/2,m_J}\right]/2$, with an $m$-independent radial Rabi frequency $\Omega_{P,r}$ and an angular matrix element $w${\color{blue}~\cite{Steck20}}. For instance, for the case of a $\sigma^+$-polarized probe $\langle e, m'_J \vert \hat\Omega_P \vert g, m_J \rangle$ $= \delta_{m'_J, m_J+1} \Omega_{P,r} \left[(w)^{l'=1,J'=3/2,m'_J}_{l=0,J=1/2,m_J}\right]$. Analogous expressions apply to the coupling-laser interaction and other polarizations. All optical couplings are diagonal in the nuclear magnetic quantum number $m_I$.

The structure of the total Hamiltonian of our ladder-type system is illustrated in Table~\ref{table1}.
It is seen that the Hamiltonian takes a block-diagonal form. The diagonal blocks contain the hyperfine (Eq.~\ref{eq:hhfs}), magnetic (Eq.~\ref{eq:hB})and field-free terms (Eq.~\ref{eq:h0}).
The decay of the system almost exclusively comes from the spontaneous emission of the 6P$_{3/2}$ level, which is included in the center diagonal block. The diagonal blocks also depend on the atom velocity, as shown in Eq.~\ref{eq:h0}.
The off-diagonal blocks contain the optical couplings (Eq.~\ref{eq:hi}), which are $m_J$-dependent according to the given polarizations.

\begin{table}[tbp]
\newcommand{\tabincell}[2]{\begin{tabular}{@{}#1@{}}#2\end{tabular}}
\renewcommand{\arraystretch}{2}
\addtolength{\tabcolsep}{5pt}
\centering
\caption{Sketch of the total Hamiltonian of our ladder-type system.}
\begin{tabular}{c c c}
$\vert r \rangle$ & $\vert e \rangle$ & $\vert g \rangle$ \\ \hline
\multicolumn{1}{|c|}{\tabincell{c}{n$S_{1/2}$\\ B}} & \multicolumn{1}{c|}{$\Omega_{C}$} & \multicolumn{1}{c|}{0} \\ \hline
\multicolumn{1}{|c|}{$\Omega_{C}$ } & \multicolumn{1}{c|}{\tabincell{c}{6$P_{3/2}$\\ B, HFS}} & \multicolumn{1}{c|}{$\Omega_{P}$} \\ \hline
\multicolumn{1}{|c|}{0} & \multicolumn{1}{c|}{$\Omega_{P}$} & \multicolumn{1}{c|}{ \tabincell{c}{6$S_{1/2}$\\ B, HFS}} \\ \hline
\end{tabular}
\label{table1}
\end{table}

As shown in Fig.~~\ref{fig1}(a) (right), the degeneracy of the Zeeman sublevels is broken when the external magnetic field is applied. The figure shows the case of a weak field (about 10~Gauss). The ground levels are well within in the linear Zeeman regime,
in which the usual $F$ and $m_F$ quantum numbers are good quantum numbers. The ground-state hyperfine Lande g-factor $g_{F,g}=1/4$.
In Fig.~1(a) the intermediate levels are still within the linear Zeeman regime, with $g_{F,e} = 2/5$, but develop significant quadratic Zeeman effect within the experimentally accessed magnetic-field range. The Rydberg state is deep in the Paschen-Back regime, where $F$ and  $m_F$ are not good quantum numbers. For $\vert r \rangle$, all magnetic sub-states have well-defined $m_J$ and the nuclear $m_I$ contributes no significant energy shift.
Hence, the Rydberg states in the magnetic field have just two sets of $m_I$-degenerate levels, which are labeled spin-up and spin-down in Fig.~\ref{fig1}(a).
While the Zeeman shifts range from the linear ($\vert g \rangle$) and the quadratic Zeeman ($\vert e \rangle$) to the deep Paschen-Back regime ($\vert r \rangle$), the quantity  $m_{I}+m_{J}$ is well-defined in all regimes.

In Fig.~\ref{fig1}(a), both probe and coupling lasers are linearly polarized in $x$-direction and consist in equal parts of $\sigma^+$- and $\sigma^-$-components with respect to the $z$-quantization axis (which is parallel to $B$, as seen in Fig.~\ref{fig1}). The major transitions are labeled A, B, C, D and A$'$, B$'$, C$'$, in analogy with the labeling used in the EIT spectra  below. In Fig.~\ref{fig1}
we indicate the Rabi frequencies $\Omega$/2$\pi$ at the beam centers, for probe power of $2~\mu$W and waist $w_0=130~\mu$m, and for coupler power of $20~$mW and waist $w_0=130~\mu$m. The transitions from intermediate states into Rydberg states are divided into strong transitions, (Type-I, green solid lines), and weak transitions (Type-II, green dashed lines). The line-strength difference results  large parts from the degree of electron spin overlap between the intermediate and Rydberg states. In the calculation of coupler Rabi frequencies account must be taken of the fact that the Rydberg levels are in the Paschen-Back regime of the HFS while the intermediate levels are in the Zeeman regime.

The EIT line positions as a function of coupling-laser detuning, $\Delta_C$, follow from the Zeeman shifts $\Delta_g$,  $\Delta_e$ and $\Delta_r$ of the ground, intermediate and Rydberg levels, and the fixed detuning of the probe laser, $\Delta_P$.
In the linear Zeeman regime, $B \lesssim 10~$Gauss, and for ground- and
intermediate-state magnetic quantum numbers $m_{F,g}$ and $m_{F,e}$, and respective hyperfine g-factors $g_{F,g}=0.25$ and $g_{F,e}=0.4$, the shifts from the field-free level positions are

\begin{eqnarray}
\Delta_{g}&=&\mu_{B} B g_{F,g} m_{F,g} \nonumber\\
\Delta_{e}&=&\mu_{B} B g_{F,e} m_{F,e} \nonumber\\
\Delta_{r}&=&\mu_{B} B (g_{J,r}m_{J}+g_{I}m_{I}) + A_{hfs,r} m_I m_J
\label{eq:deltas}
\end{eqnarray}

For the 33S$_{1/2}$ Rydberg level of $^{133}$Cs, the hyperfine coupling constant $A_{hfs,r}$ is only about $0.5~$MHz. Hence, the Rydberg level is deep in the Paschen-Back regime of the HFS, and there are no apparent Rydberg HFS effects. (The Rydberg hyperfine term $A_{hfs,r} m_I m_J$ is still included in our calculations). Moreover, $g_{J,r} \approx 2.00232$ (the electron $g$-factor) and $g_I \sim -10^{-4} g_{J,r}$  ({\sl{i.e.}}, the Rydberg-level shifts due to the nuclear magnetic moment are even less than those due to the Rydberg-HFS). As a result, there are only two relevant Rydberg energy levels, $m_J = \pm 1/2$, and they are essentially $m_I$-degenerate (see Fig.~1~(a)).
Unlike in many other EIT scenarios, all three levels shift as a function of $B$, and there is a large number of electric-dipole-coupled transitions with different detunings (see Fig.~1~(a)). For fields $B \gtrsim 10~$Gauss the second-order Zeeman effect of the intermediate level become important; in this case the magnetic level shifts $\Delta_g$ and  $\Delta_e$ are obtained by diagonalization of the optical-field-free Hamiltonian in the $\{ |m_I, m_J \rangle \}$ basis.

The EIT line positions for fixed probe detuning $\Delta_P$ are obtained by requiring resonance for both the probe and the coupler laser on a three-level ladder connected via dipole-allowed transitions. The resonance conditions are solved by treating the atom velocity $v$ and coupling-laser detuning $\Delta_C$ as free variables. The solutions are
\begin{eqnarray}
v&=&  \frac{\lambda_P}{2 \pi} (\Delta_P + \Delta_g - \Delta_e)  \nonumber\\
\Delta_{C}&=& \Delta_r + \Delta_e \left( \frac{\lambda_P}{\lambda_C}-1 \right) - (\Delta_g + \Delta_P) \frac{\lambda_P}{\lambda_C} \nonumber\\
~ &=& \Delta_r + 0.67 \Delta_e - 1.67 (\Delta_g + \Delta_P)\quad,
\label{eq:EITres}
\end{eqnarray}
with sets of detunings $\Delta_g$, $\Delta_e$ and $\Delta_r$ from Eq.~\ref{eq:deltas}, that correspond with sets of three states that have dipole-allowed probe and coupling transitions. There are many solutions, as indicated in Fig.~1(a).

\begin{figure*}[t]
\centering
\includegraphics[width=0.7\textwidth]{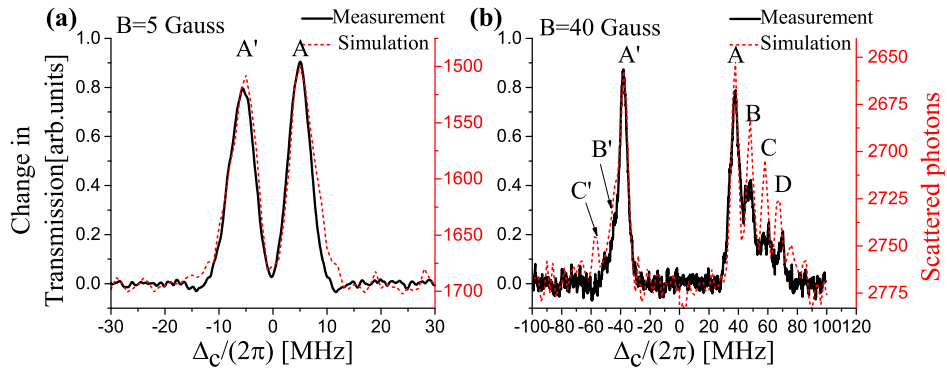}
\caption{ (Color online) Measurements (left axes, black solid lines) and simulations (right axes, red dashed lines) of Rydberg EIT spectra with magnetic fields of 5~Gauss (a) and 40~Gauss (b). The probe and coupling beams are both linearly polarized in $x$-direction (transverse to the magnetic field). The experimental data show increase in transmission above its coupler-laser-free value. EIT on the various three-level ladder systems identified in Fig.~1~(a) results in several peaks. The QMCWF simulations show the number of photons scattered by a typical atom sample in the cell (note the number of scattered photons is plotted in descending direction). EIT corresponds with a reduction in photon scattering. With increasing magnetic field, the Rydberg EIT line splits into two approximately symmetric main peaks A$'$ and A. Asymmetric satellite lines, labeled B$'$, C$'$ and B, C, D, appear outside of the interval between A$'$ and A. The peak labels correspond to the transition labels in Fig.~1~(a).}
\label{fig2}
\end{figure*}

\subsection{Monte-Carlo approach to model optical pumping and EIT lineshapes}

To interpret the rich structure that we measure in our experimental Rydberg EIT spectra in magnetic fields (multiple line positions, line strengths and line shapes), it is essential to employ a quantitative model that accounts for the exact dependence of the optical couplings on magnetic quantum numbers and beam polarizations, atom decay, optical pumping among the magnetic sub-levels, the nonlinear Zeeman effect, and Rydberg-level dephasing. In the geometry given in Fig.~1~(b), in general all 6$S_{1/2}$, 6$P_{3/2}$ and $nS_{1/2}$-states are coupled to each other by the optical fields and the spontaneous decay of the excited states. For $^{133}$Cs, which has $I=3.5$, the wavefunction dimension therefore is $N = 64$. This is sufficiently large to seek the benefits of the quantum Monte-Carlo wavefunction (QMCWF) method to determine the density matrix, rather than to directly solve the quantum Master equation.
In the QMCWF approach~{\color{blue}\cite{Castin22,Dalibard23}}, one generates large sets of quantum trajectories whose evolution consists of deterministic segments of Hamiltonian propagation (with a non-Hermitian effective Hamiltonian) and discrete, stochastic quantum jumps. After each quantum jump the wave-function is re-set to a new, normalized wavefunction that is consistent with the quantum measurement of a spontaneously emitted photon. The density matrix is constructed from an average over a sufficiently large sample of quantum trajectories. It has been shown that the QMCWF method is equivalent to the standard density matrix analysis~{\color{blue}\cite{Castin22,Dalibard23}}, an instance of which is found in our recent work~{\color{blue}\cite{Bao15}}. If the dimension of the quantum system is large, as in the present case with $N=64$, the number of variables in the wave-function process ($\sim N=64$) is so much smaller than the number of variables in the density-matrix analysis ($\sim N^{2} = 4096$) that the QMCWF becomes numerically a better choice. The QMCWF method also scales more readily to even larger systems, such as ones that involve D-state Rydberg levels (for which it would be $N=128$). Therefore, for our interpretation of the $^{133}$Cs Rydberg EIT spectra we employ the QMCWF method. The fundamentals of the method are explained in Refs.~{\color{blue}\cite{Castin22,Dalibard23}}; our own work on QMCWF in context with laser cooling in a variety of optical lattices is summarized in Ref.~{\color{blue}~\cite{Morrow24}} and references therein. In the present work, we use the QMCWF to obtain the probe-photon scattering rate per atom on a grid of atom velocities $v$ in the cell and over a range of coupling-laser detunings $\Delta_C$, for selected values of the magnetic field $B$ and radial Rabi frequencies, $\Omega_{P,r}$ and $\Omega_{C,r}$, and for the laser polarizations we have investigated in the experiment. As sufficiently large atom samples we typically use 2000 quantum trajectories integrated over 1~$\mu$s each.

\section{Experimental spectra of Rydberg EIT in magnetic field}

\subsection{Magnetic-field-induced EIT line asymmetry}

Figure~2 shows experimental Rydberg EIT spectra in axial magnetic fields of $B=5$~Gauss and 40~Gauss in  (a) and (b), as well as the results of corresponding QMCWF simulations. The probe and coupling beams are both linearly polarized in $x$-direction. The intrinsic EIT-linewidth is $\sim 8$~MHz; it is mostly due to power broadening on the probe and coupler transitions as well as residual laser linewidth effects. In the 5-Gauss case, it is observed that the Rydberg EIT spectrum splits symmetrically into two peaks of approximately equal shape. The corresponding EIT resonance conditions that follow from Eq.~\ref{eq:EITres} correspond to the cascade $\vert 6\emph{S}_{1/2} F_g=4, m_{F,g} = 4 \rangle \rightarrow$ $\vert 6\emph{P}_{3/2} F_e=5, m_{F,e}=5 \rangle \rightarrow$ $\vert 33\emph{S}_{1/2}, m_{J,r}=1/2, m_{I}=3.5 \rangle$, labeled A (blue-shifted), and the cascade $\vert 6\emph{S}_{1/2} F_g=4, m_{F,g} = -4 \rangle \rightarrow$ $\vert 6\emph{P}_{3/2} F_e=5, m_{F,e}=-5 \rangle \rightarrow$ $\vert 33\emph{S}_{1/2}, m_{J,r}=-1/2, m_{I}=-3.5 \rangle$, labeled A' (red-shifted). The velocities at which the EIT resonances occur are $-6.0$~m/s (blue-shifted peak, A) and $6.0$~m/s (red-shifted peak, A'). For the velocity class near -6.0~m/s, the $\sigma^+$-component of the $x$-polarized probe field is closer to resonance, leading to optical pumping close the state $m_{F,g} =4$, which in turn produces the strong blue-shifted EIT peak A, in conjunction with the $\sigma^-$-component of the $x$-polarized coupling field. Similarly, for $v \approx +6.0$~m/s the $\sigma^-$-component of the probe field is closer to resonance, leading to optical pumping close to $m_{F,g} =-4$, which produces the red-shifted peak A', in conjunction with the $\sigma^+$-component of the coupler field. Since the optical pumping is not perfect, a small fraction of the atoms reside in $m_{F,g} = \pm 3, \pm 2, ...$. Those atoms are driven on the B and B' etc. sequences (see Fig.~1~(a)), leading to a slight broadening on the blue side of A and the red side of A'.

Without quadratic Zeeman shift, and if the $F_{e} = 4, 3$ HFS levels were several GHz away from $F_{e} = 5$, the situation in Fig.~2~(a) would exhibit perfect symmetry about  $\Delta_C=0$. The slight asymmetry Fig.~2~(a), seen in both the experiment and the simulation, is a first indication that quadratic Zeeman shifts, and possibly other intermediate-state HFS levels, play a role even at small fields. It is somewhat surprising that at fields as small as 5~Gauss there already is evidence for  the effects of quadratic Zeeman shifts on EIT spectra.

In the 40-Gauss data, shown in Fig.~2(b), the weak satellite peaks, labeled B$'$, C$'$ and B, C, D, are largely resolved and appear outside the interval between A$'$ and A. The satellite peaks are cascades identified by corresponding labels in Fig.~1(a). Both the separations and the strengths of the satellite peaks are  highly asymmetric in the 40-Gauss magnetic field. The asymmetry in  the positions and the strengths of the satellite peaks increases with magnetic field. It is due to the quadratic Zeeman effect, which introduces the asymmetry in the line splitting and, more importantly, in the optical-pumping efficiency and the resultant relative strengths of the satellite peaks relative to the main (A and A$'$) peaks (see Sec.~\ref{subsec:opu}).

\begin{figure}[b]
\includegraphics[width=0.4\textwidth]{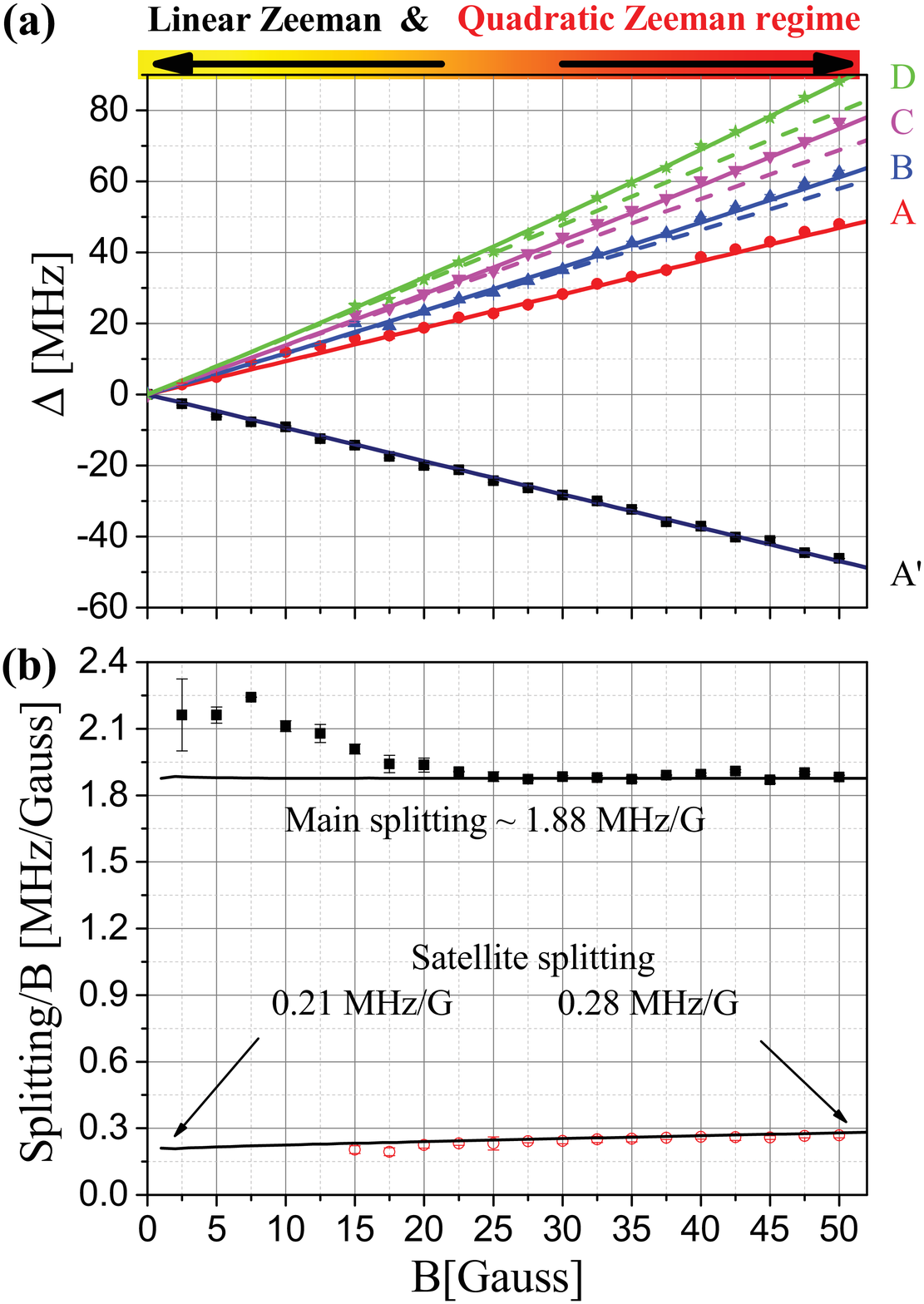}
\caption{(Color online) (a) Experimental (symbols) and calculated (solid lines) frequency shifts of main peaks A$'$ (black squares) and A (red circles), and of the blue-shifted satellite peaks B (blue up triangles), C (pink down triangles) and D (green stars).
The dashed lines show line shifts calculated under the assumption of no quadratic Zeeman effect.
(b)~Line intervals divided by magnetic field for the A'-A interval (black squares), and for the average separations between adjacent lines pairs, A-B, B-C and C-D (red circles), as a function of magnetic field. Symbols show experimental data, solid lines show calculated values that take quadratic Zeeman shifts into account.}
\label{fig3}
\end{figure}

\subsection{EIT line positions and splittings}
\label{subsec:opu}

We have taken a series of data similar to Fig.~2 for a series of magnetic fields.
The measured and calculated shifts of the two main EIT peaks, A and A', and of the three blue-shifted satellite peaks, B, C and D, are shown as a function of magnetic field in Fig.~3~(a). In the calculations we use Eq.~\ref{eq:EITres} with level shifts
$\Delta_g(B)$, $\Delta_e(B)$ and $\Delta_r(B)$ taken from numerical diagonalization of $\hat{H}_{hfs} + \hat{H}_{B}$ (see Eqs.~\ref{eq:hhfs} and~\ref{eq:hB}) within the respective ground-, intermediate- and Rydberg-level subspaces. The larger error bars and the absence of satellite peaks in magnetic fields $<15$~Gauss reflect the fact that in such small fields the Zeeman splittings between the satellite lines cannot be experimentally resolved. In Fig.~3~(a) it is seen that the two main peaks shift linearly as a function of magnetic field. This is expected because these peaks only involve aligned states, {\sl{i.e.}} states with $m_I + m_J = I+J$. Those states have no quadratic Zeeman effect in the magnetic-field range of interest. Close inspection of Fig.~3~(a) further reveals that the satellite peaks deviate from the dashed straight lines, which mark the line positions one would observe under absence of quadratic Zeeman shifts. The experimentally observed and calculated deviations of the EIT peaks from the straight lines are in excellent agreement. They are almost entirely due to the quadratic Zeeman effect of the magnetic sublevels of the intermediate (6P$_{3/2}$) state, and they become significant at fields larger than about 20~Gauss. At the largest field in Fig.~3 the nonlinear contribution to the shift of the D-peak approaches 10~MHz, corresponding to about ten times the uncertainty in the experimental line positions.

Fig.~3~(b) shows measured and calculated intervals between the two main peaks divided by the magnetic field, $\Delta \chi_{AA'} = (A - A')/B$, as a function of the line positions $A$ and $A'$ and the magnetic field $B$.
The calculated value for $\Delta \chi_{AA'}$ is fixed at 1.88~MHz/Gauss, while the observed $\Delta \chi_{AA'}$-value approaches the calculated result to within the uncertainty for fields $B$ exceeding about 15~Gauss. The disagreement between the calculated and observed $\Delta \chi_{AA'}$-values seen at small fields likely is due to the fact that in small fields the satellite and main lines are not resolved. In that case, the A, B, C and D lines merge into a single line whose center is blue-shifted relative to the calculated A-line. Likewise, the A', B' etc. lines merge into a single line that is red-shifted relative to the calculated A'-line. As a result, at small fields the measured $\Delta \chi_{AA'}$-values exceed the calculated ones.

In Fig.~3(b) we also show the average splitting between adjacent satellite peaks divided by $B$, denoted $\Delta \chi_{Sat}$. The measured $\Delta \chi_{sat}$-value gradually increases from about 0.25~MHz/Gauss at 15~Gauss, where the satellite lines start to become resolved, to 0.28~MHz/Gauss at 50~Gauss. The calculated $\Delta \chi_{sat}$-value increases from 0.21~MHz/Gauss at zero field to 0.28~MHz/Gauss at 50~Gauss; the increase is due to the quadratic Zeeman shift. Considering the EIT linewidth in the experiment, experimental and calculated data in Fig.~3~(b) are in good agreement and validate the importance of the quadratic Zeeman effect even in small magnetic fields.

\subsection{Interplay between optical pumping and EIT}
\label{subsec:opu}

\begin{figure*}[t]
\includegraphics[width=1.1\textwidth]{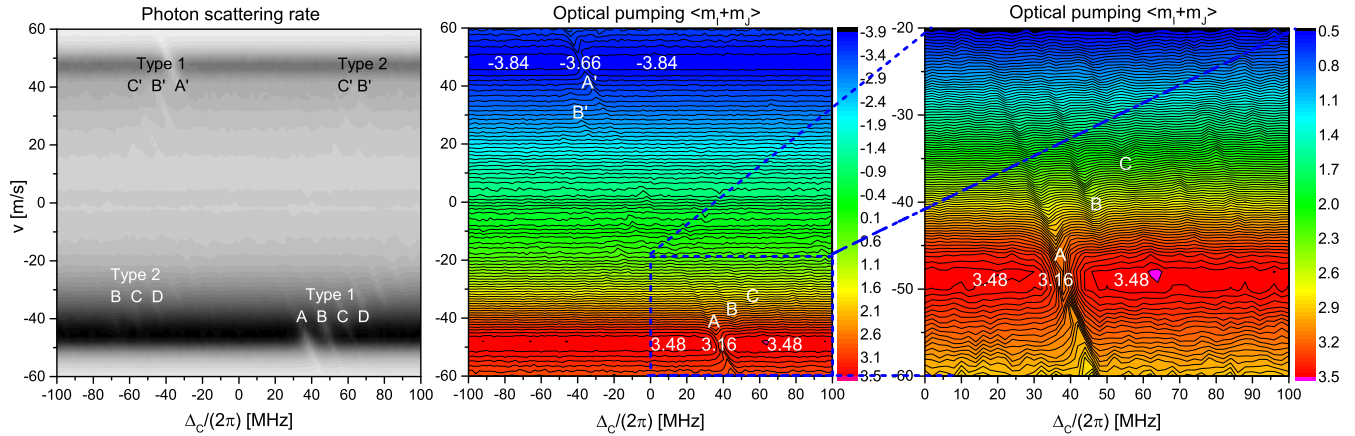}
\caption{(Color online) Left: simulated photon scattering rate vs coupler detuning and atom velocity for 40~Gauss magnetic field in $x$-polarized  coupler and probe fields with respective radial Rabi frequencies $\Omega_{C,r}/(2 \pi)=8$~MHz and $\Omega_{P,r}/(2 \pi)=2$~MHz, on a linear gray scale ranging from 0 (white) to $2.8 \times 10^6$~s$^{-1}$  (black) per atom. The labels of the features visible in the plot correspond with the labels Fig.~1. Middle and right (zoom-in): simulated expectation values for the ground-state $\langle m_{I} + m_{J} \rangle$, obtained from the same simulation as on the left, on color scales given by the color bars. Regions of strong photon scattering correspond with fairly efficient optical pumping into the aligned states, $\vert m_{F,g} = \pm 4 \rangle$.
In the regions of large photon scattering the Zeeman shifts of the probe cycling transitions,
$\vert m_{F,g} = \pm 4 \rangle$ $\rightarrow \vert m_{F,e} = \pm 5 \rangle$, are compensated by the Doppler effect at
$v \approx \pm$48~m/s. The numbers indicated on the figures show the approximate values of $\langle m_{I} + m_{J} \rangle$ on the main EIT resonances, $A$ and $A'$, and close to them. It is seen that the EIT-induced reduction in probe photon scattering is accompanied by a reduction in optical-pumping efficiency. The plots also show a significant asymmetry between  positive and negative $v$, and between positive and negative $\Delta_C$.}
\label{fig4}
\end{figure*}

In this section we discuss optical-pumping effects and their relation with EIT. In optical pumping~{\color{blue}\cite{Happer25}}, the distribution of atomic populations over the Zeeman sublevels becomes non-thermal after several absorption-emission cycles.
Without magnetic field and for zero-velocity atoms, $x$-polarized probe light optically pumps atoms mostly into the states $|6S_{1/2}, F=4, m_{F,g}=\pm 4 \rangle$, implying a symmetric situation in which the A and A' probe transitions in Fig.~1 account for most absorption. The  optical pumping is reduced when the degeneracy of the Zeeman transitions is broken.
Under presence of a magnetic field pointing along $z$, the transitions driven by the $\sigma^+$- and $\sigma^-$-components of the $x$-polarized probe field become detuned from one another, due to the different $g_F$-factors of the $6S_{1/2}$ and $6P_{3/2}$ states. In this case, optical pumping primarily occurs for combinations of atom velocities and probe detunings for which either the A or the A' probe transition is near-resonant (but not both). This is illustrated by simulation results in Fig.~\ref{fig4}, which is for a magnetic field of 40~Gauss and $\Delta_P=0$. As seen from Eq.~\ref{eq:EITres}, for fixed $\Delta_P$ the free parameters available to achieve the EIT resonance condition on both the probe and the coupler transitions are the atom velocity $v$ in $z$-direction and the coupler detuning $\Delta_C$. The range of available velocity classes to satisfy the conditions follow from the Maxwellian density distribution of thermal atoms, $\emph{N}(v)=\frac{\emph{N}_{0}}{u\sqrt{\pi}} e^{-v^{2}/u^{2}}$ , where $u/\sqrt{2}$  is the root-mean-square atomic velocity in one dimension. Here, $u/\sqrt{2} \approx 140$~m/s, $\emph{N}_{0}$ denotes the total atomic volume density in the cell, and $\emph{N}(v)$ is the velocity-dependent volume density per  m/s. According to Eq.~\ref{eq:EITres}, the resonant velocities only depend on $\Delta_P$, which is zero in the simulation, and on the ground- and intermediate-level Zeeman shifts (notably not on $\Delta_C$).

In Fig.~\ref{fig4}, the A-transition is near-resonant for atoms with velocities  $\approx -48$~m/s. This leads to optical pumping to near $\langle m_{I} + m_{J} \rangle = 4$, as shown in the middle and right panels in Fig.~\ref{fig4}, as well as high photon scattering rates for atoms in this velocity class, as seen in the left panel in Fig.~\ref{fig4}. Similarly, the A'-transition is near-resonant for atoms with velocities $\approx 48$~m/s, leading to optical pumping to near $\langle m_{I} + m_{J} \rangle = -4$ and high photon scattering rates for atoms in that velocity class. The left panel in Fig.~\ref{fig4} further shows that atoms near those velocities account for most absorption, independent of $\Delta_C$. Atoms with velocities between $-48$~m/s and 48~m/s are optically pumped to some intermediate value of  $\langle m_{I} + m_{J} \rangle$; however, these atoms have low photon scattering rates, and they contribute little to the overall absorption behavior. Also, in our magnetic-field range the resonant velocities are much smaller than the root-mean-square atomic velocity. Therefore, the density reduction in $\emph{N}(v)$ due to the Maxwell distribution is only on the order of $5\%$ and has no significant effect on the spectra.
\begin{figure*}[t]
\centering
\includegraphics[width=0.7\textwidth]{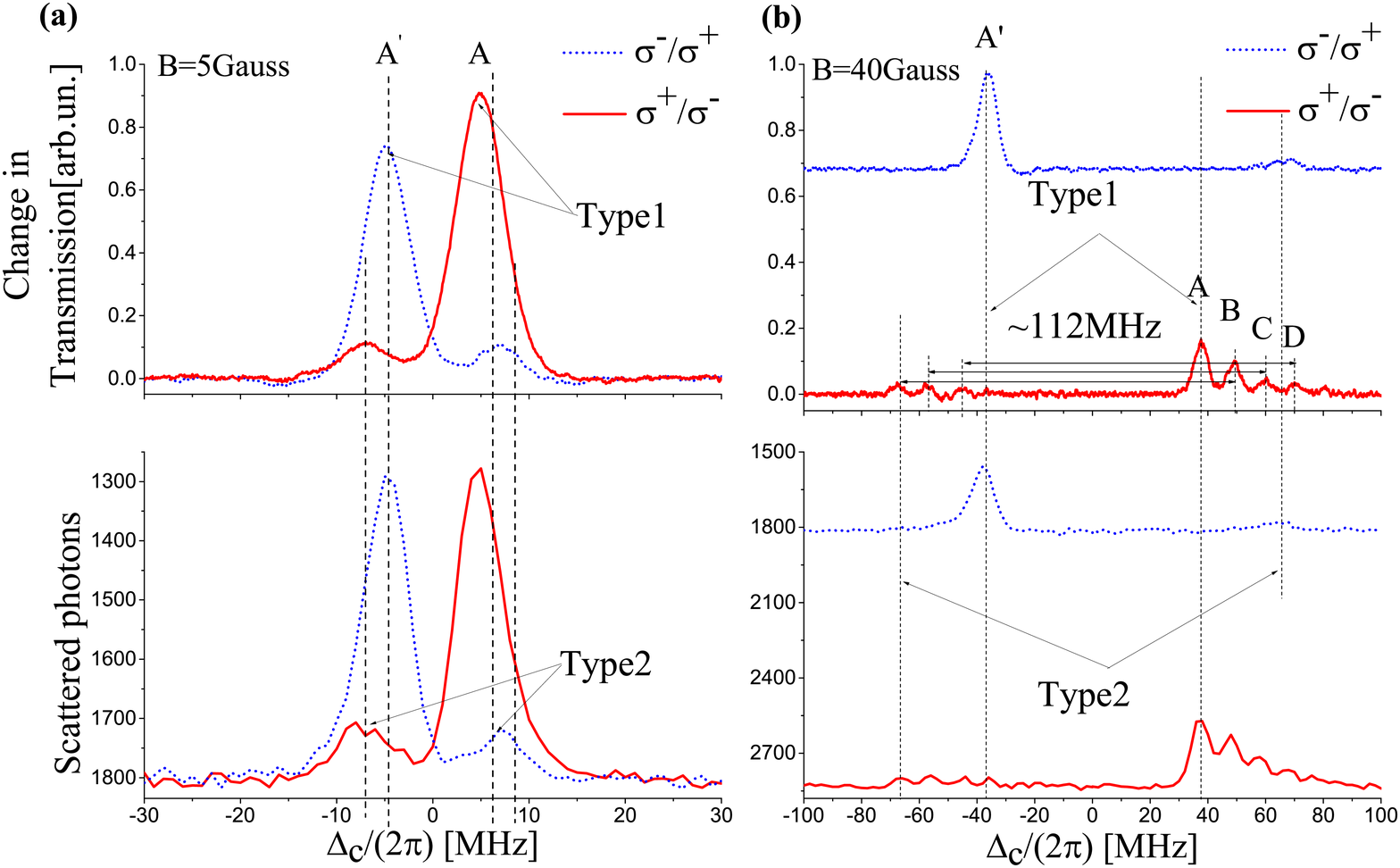}
\caption{(Color online) Experimental (top) and calculated (bottom) Rydberg EIT spectra in 852~$\sigma^{+}$/ 510~$\sigma^{-}$ and 852~$\sigma^{-}$/ 510~$\sigma^{+}$ polarized probe and coupler fields. The magnetic fields are 5~Gauss (a) and 40~Gauss (b). The EIT spectra in (a) are approximately symmetric about $\Delta_C=0$, while they are highly asymmetric in shape and background absorption in (b). The asymmetry is due to the quadratic Zeeman effect of the intermediate 6$P_{3/2}$-level. The line labels and types are explained in Fig.~\ref{fig1} and in the text.}
\label{fig5}
\end{figure*}

Further inspection of Fig.~\ref{fig4}  shows an asymmetry between positive and negative velocities, as well as in the degree to which the  $\langle m_{I} + m_{J} \rangle$-values approach the limits $\pm 4$ (corresponding to perfect optical pumping into the aligned states, which have $m_{I} + m_{J} = \pm (I+J)$). The asymmetry is a result of the quadratic Zeeman effect in the $6P_{3/2}$ state. The levels $m_{I} + m_{J} = 5,4,3 ...$ are considerably more closely spaced than the levels $m_{I} + m_{J} = -5,-4,-3 ...$ (see, for instance, Ref.{\color{blue}~\cite{Steck20})}.
Hence, the $m_{I} + m_{J} = 5,4,3 ...$ Zeeman splittings in $6P_{3/2}$ match the splittings of the $m_{I} + m_{J} = 4,3,2 ...$ in the ground state $6S_{1/2}$ much more closely than the $m_{I} + m_{J} = -5,-4,-3 ...$ splittings in $6P_{3/2}$ match the splittings of the $m_{I} + m_{J} = -4,-3,-2 ...$ in $6S_{1/2}$. This  leads to higher photon scattering rates at $v\approx -48$~m/s.

In the middle and right panels of Fig.~\ref{fig4} we indicate the extremal values of $\langle m_{I} + m_{J} \rangle$. At coupler detunings away from the EIT features, the $\langle m_{I} + m_{J} \rangle$-values peak at -3.84 (lower photon scattering, $v \approx 48$~m/s) and 3.48 (higher photon scattering, $v \approx -48$~m/s). It is therefore seen that higher scattering rates correspond with somewhat less efficient pumping. In the EIT signals, the less efficient optical pumping at $v \approx -48$~m/s leads to quite pronounced satellite peaks B, C and D on the blue side of peak A, at $\Delta_C \sim 50$~MHz. From Figures~\ref{fig1}-\ref{fig4} it is evident that the peak A corresponds with $m_{F, g}=4$, B corresponds with $m_{F, g}=3$ etc. Hence, the incomplete optical pumping into $m_{F, g}=4$ causes the relatively strong B, C and D satellite peaks. Conversely, the more efficient optical pumping at $v \approx +48$~m/s into  $m_{F, g}=-4$ leads to less pronounced satellite peaks B' and C' on the red side of peak A', at $\Delta_C \sim -50$~MHz in the EIT spectrum.

It is further noteworthy that away from the EIT resonances the coupler beam has very little effect on optical-pumping behavior. This reflects the fact that away from the EIT resonances there is no velocity-detuning pair $(v, \Delta_C)$ that satisfies a EIT double-resonance for both coupler and probe fields. However, on the EIT resonances the coupler inhibits probe-photon scattering, leading to substantial drops in photon scattering and optical-pumping  efficiency. For instance, the EIT peak labeled A reduces $\langle m_{I} + m_{J} \rangle$ from about 3.48 to 3.16, and the EIT peak labeled A' reduces $\vert\langle m_{I} + m_{J} \rangle \vert$ from about 3.84 to 3.66. The corresponding ``grooves'' in the photon-scattering and optical-pumping maps are clearly seen in Fig.~\ref{fig4}. We have found the asymmetry trends in Fig.~\ref{fig4} as well as the back-action of EIT onto optical pumping at other magnetic fields as well. The asymmetries in photon scattering and optical pumping, seen between positive and negative coupler detunings $\Delta_C$ and velocities $v$, are due to the nonlinear Zeeman effect and generally increase with magnetic field. Asymmetry already becomes evident at fields as small as about 5~Gauss; in lesser fields the nonlinear Zeeman effect is so small that spectra, photon-scattering and optical-pumping maps are essentially symmetric in $\Delta_C$ and $v$.

\section{Optical-pumping effects on EIT for other polarizations}
\label{sec:pols}

We have also observed asymmetry of Rydberg EIT spectra in a series of analogous experiments for polarization cases 852~$\sigma^{+}$/ 510~$\sigma^{-}$ and 852 $\sigma^{-}$/ 510 $\sigma^{+}$. The experimental results and corresponding simulations are shown in top and bottom panels in Fig.~\ref{fig5}, respectively. When the magnetic field is only 5~Gauss, the Rydberg EIT spectra are almost symmetric about $\Delta_C$ because in fields that small there is no significant quadratic Zeeman effect. In contrast, at 40~Gauss the Rydberg EIT spectra for the two polarization cases in Fig.~\ref{fig5} are highly asymmetric, in a way that closely resembles the asymmetry seen in Fig.~\ref{fig2}. Especially, there are three clear satellite peaks, B, C and D, on the blue side of A, while there are no significant satellite peaks, B' etc, on the red side of A'. The increase in asymmetry at larger magnetic fields reflects the increasing importance of the quadratic Zeeman effect in those fields.

It is also noted that the quadratic Zeeman effect brings about an asymmetry in overall background absorption. At 5~Gauss the absorption away from the EIT peaks is almost the same for $\sigma^{+}$- and $\sigma^{-}$-polarized probe fields, while at 40~Gauss the $\sigma^+$-polarized probe is considerably more absorbed than the $\sigma^-$-polarized probe. Note the corresponding difference in the number of scattered photons in the simulations (bottom panels in Fig.~\ref{fig5}). Another interesting observation in the 40-Gauss data in Fig.~\ref{fig5} is the clear appearance of the ``Type-II''-peaks. As seen in Fig.~\ref{fig1}(a), the ``Type-II''-transitions, shown by dashed green arrows, have much lower Rabi frequencies than the dominant ``Type-I''-transitions (solid green arrows). The difference in Rabi frequencies is caused by different degrees of electron-spin overlap between the intermediate and the Rydberg states. Since the Rydberg state
is in the deep hyperfine Paschen-Back regime, the frequency shift between corresponding Type-I and Type-II lines is $B \times 2.8$~MHz/Gauss, corresponding to 112~MHz in the right panels in Fig.~\ref{fig5} (see horizontal arrows).

\section{CONCLUSION}

We have performed Rydberg EIT spectroscopy in a cesium room-temperature vapor cell when an axial magnetic field in the range between 0 and 50~Gauss is applied. The Rydberg EIT spectra exhibit characteristic sets of peaks and
satellite lines that become highly asymmetric, in fields as small as $\sim 40$~Gauss.
We have modeled the spectra using Quantum Monte Carlo wave-function simulations. The EIT spectra were found to be quite susceptible to the quadratic Zeeman effect, which leads to considerable asymmetry in both the line splittings and the line strengths and shapes of the red-detuned versus the blue-detuned Zeeman features of the EIT lines.  The simulations also showed that the asymmetry in EIT satellite line strengths is, in large parts, due to differences in the optical-pumping efficiency on the red versus the blue-shifted parts of the spectra. The presented results are a prerequisite for spectroscopic, EIT-based electric-field diagnostics that involve a background magnetic field, such as studies of magnetized-plasma physics in cesium or rubidium cells.

\section{ACKNOWLEDGEMENTS}
The work was supported by the National Natural Science Foundation of China (Grants No.61378013, No. 91536110, No. 61475090, No. 61505099, No. 61675123), G.R. acknowledges support from the NSF (Grant No. PHY-1506093) and the BAIREN plan of Shanxi province.

\end{document}